\documentclass[aps,prApplied,reprint,superscriptaddress]{revtex4-2}

\usepackage{dcolumn}
\usepackage{graphicx}
\usepackage{subfigure}
\usepackage{textcomp}
\graphicspath{{figures/}}

\begin{document}
	
	
	\title{A Magnetic Compression method for sub-THz electron beam generation from RF freqencies}
	
	
	\author{An Li}
	\affiliation{Department of Engineering Physics, Tsinghua University, Beijing 100084, People’s Republic of China}
	
	\author{Jiaru Shi}
	\affiliation{Department of Engineering Physics, Tsinghua University, Beijing 100084, People’s Republic of China}

	\author{Hao Zha}
	\email{ZhaH@mail.tsinghua.edu.cn}
	\affiliation{Department of Engineering Physics, Tsinghua University, Beijing 100084, People’s Republic of China}
	
	\author{Qiang Gao}
	\affiliation{Department of Engineering Physics, Tsinghua University, Beijing 100084, People’s Republic of China}

	\author{Huaibi Chen}
	\affiliation{Department of Engineering Physics, Tsinghua University, Beijing 100084, People’s Republic of China}
	
	
	\date{\today}
	
	\begin{abstract}
		
		Current THz electron sources struggle with low energy gain and device miniaturization. We propose a magnetic compression method designed for relativistic electrons to perform post-compression on the beam from radiofrequency accelerators,  to produce sub-THz electron beam with exceptionally high energy ($>1$ J).	Through simulation studies, we longitudinally compress  a relativistic electron beam with energy of 60 MeV and frequency of 3 GHz across a time span of 24 ns, yielding an electron pulse train at a 0.1 THz. The compressed beam exhibits a pulse width of 0.8 ns, a total charge of 24 nC, and an energy of 1.4 J,   providing a new potential for ultra-high-energy THz electron beams generation.

	\end{abstract}
	
	
	\maketitle
	
	
	\section{Introduction}
	
	THz sources based on relativistic electron beams can generate high-intensity THz radiation and achieve continuous tuning of frequency through compression of the electron beam. These sources hold significant potential for applications in various scientific fields such as biomedicine\cite{yang2016biomedical,sun2017recent}, materials science\cite{Material1,Material2,Material3}, astronomy\cite{kulesa2011terahertz,siegel2007thz}, among others.
	
	Current high-gradient THz electron acceleration schemes, such as Laser Wakefield Acceleration (LWFA)\cite{tajima1979laser}, Dielectric Laser Acceleration (DLA)\cite{RevModPhys.86.1337, black2019net}, THz electron acceleration, etc, offer compact solutions for achieving relativistic THz electron beam generation. However, the accelerated beam charge and energy gain remain low\cite{hibberd2020acceleration,xu2021cascaded,zhao2020femtosecond}, and the miniaturization of devices also poses challenges in fabrication.
	
	If it is possible to longitudinally compress the electron beam generated by traditional RF acceleration structures on a large time scale, increasing its frequency from the radio freqency significantly to the THz range, there is an opportunity to generate THz electron beams with a dramatically large charge.
	
	However, the issue lies in the fact that due to the high speed of relativistic electron beams, which is very close to the speed of light, compression methods based on velocity modulation are typically inefficient\cite{VelocityBunching,MagnetCompress,CoulombCompress}. For example, a chicane structure on a scale of 10 meters can only achieve compression on the order of picoseconds.
	
	To address this issue, we propose a relatively compact magnetic compression method for relativistic electron macro-pulses, which can efficiently convert energy modulation into density modulation. It has the potential to generate THz electron beams with a significantly high intensity using combination of RF accelerator and post compression set-up.

	In this study, we employed CST for magnet design and electro-magnetic field simulation. GPT and MATLAB are employed for particle dynamics simulation and transport system analyzation, validating the compression methodology.

	\section{Compressing method}
	
\begin{figure}[htbp]
		\includegraphics[width=8.5cm]{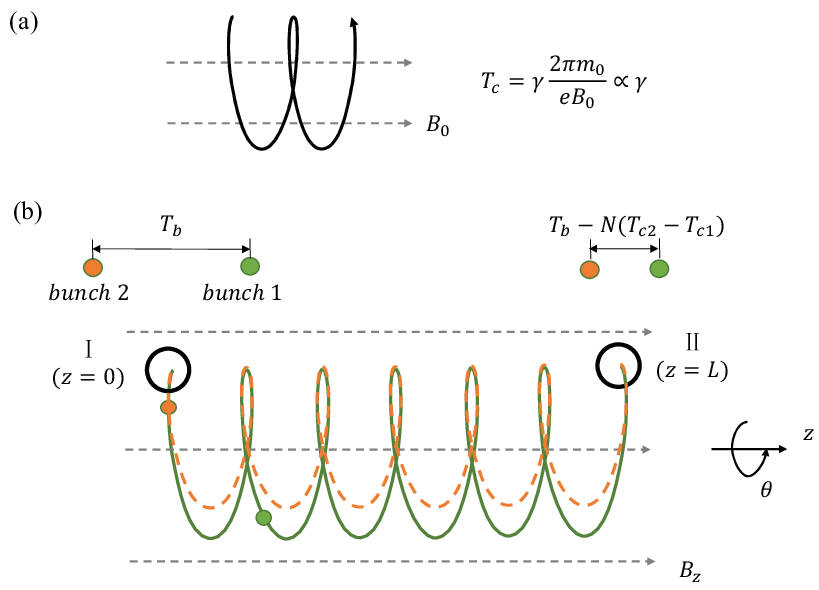}
		\caption{(a)Spiral movement of a electron in a uniform magnetic field $B_0$ (b) Compressing principle}
		\label{FIG.1}
	\end{figure}

	As shown in FIG.\ref{FIG.1}(a), in a uniform magnetic field, when the velocity of an electron makes a non-90° angle with the field vector, the electron spirals along the field direction. 

	Meanwhile, the electron's cyclotron period $T_c$ is independent of the spiral angle and always proportional to its energy $\gamma$.
	
	\begin{equation}
		\label{eq1}
		T_c=\gamma \frac{2 \pi m_0}{eB_0} \propto \gamma
	\end{equation}

	Where $m_0$ denotes the rest mass of the electron, and $B_0$ represents the magnitude of the uniform field.

	This provides a great concept for achieving density modulation of relativistic electron beams: although we cannot induce significant differences in electron velocity through energy modulation, we can leverage the dispersion effect in this simple uniform field to create noticeable differences in trajectory length, thereby enabling compression. This is somewhat akin to the effect of the bending magnet section in a chicane structure, but due to the long period of spiral motion, there exists a momentum compaction factor $R_{56}$ significantly larger than that of a chicane structure.
	
	Let's assume that the electrons complete $N$ cycles of spiral motion in the ideal uniform field, as detailed and visualized in FIG. \ref{FIG.1}(b). $R_{56}$ of the transmission process can be derived as:
	
	\begin{equation}
		\label{eq2}
		R_{56} = \frac{\Delta s}{\delta} = \beta c NT_c
	\end{equation}

	More specifically, BUNCH 1 and BUNCH 2 is initially separated by a time interval $T_b$ at position I. $\gamma_1 > \gamma_2$. After traversing the drift region, at position II, the separation is diminished to $T_b' = T_b - \gamma NT_c$.

	In addition to the larger coupling factor, this compression method has another advantage. The characteristics of electronic spiral motion allow us to fold a long trajectory into a smaller cylindrical space, making the device more compact and cost-effective. Compared to macro-pulse compression devices in CLIC and GELINA, it offers higher spatial efficiency.
	
	Of course, constrained by the magnetic field strength inside the solenoid cavity, for electrons of different energies and different compression time scales, we need devices of different sizes to match corresponding application requirements.
	
	For instance, a solenoid with a 1 meter outer diameter and 5 m height, operating at a current density of about $7.9 \times 10^4$ A/m, generates a 0.1 T uniform field inside its cavity. The electrons with 6.5 MeV energy will undergo spiral motion with a period of 5.4 ns. if the electrons undergo ten cycles of spiral motion from injection to extraction, momentum compaction factor of the system  $R_{56} = 16.2$ m.
	
	However, due to low reference electron energy and a limited 10\% energy dissipation acceptance range, the beam pulse width and charge are insufficient. The system can only compress beams with a total energy level in the range of 10 mJ to sub-THz frequencies.
	
	If we scale up the system size, magnetic field strength, and electron beam energy proportionally, we can greatly boost the energy of the compressed beam. This enhances the dynamic aperture and energy acceptance range while mitigating space charge effects. Here, we introduce a large-scale device enabling THz compression of 1 J beams and analyze beam characteristics pre and post-compression through dynamic simulations.

	\subsection{compressing set-up}
	
	In this study, we utilized a device measuring 6 meters in outer diameter and 29 meters in height to facilitate large-scale compression of electron beams within the energy range of 60 MeV ± 10\%. This compression enables an increase in electron pulse train repetition frequencies from 3 GHz to 0.1 THz. Moreover,  it enables the extraction of electron beams with a total charge of 24 nC, resulting in the generation of a THz electron pulse train with energy surpassing 1.4 J over a duration of 0.8 ns.
	
	\begin{figure}[htbp]
		\includegraphics[width=8.5cm]{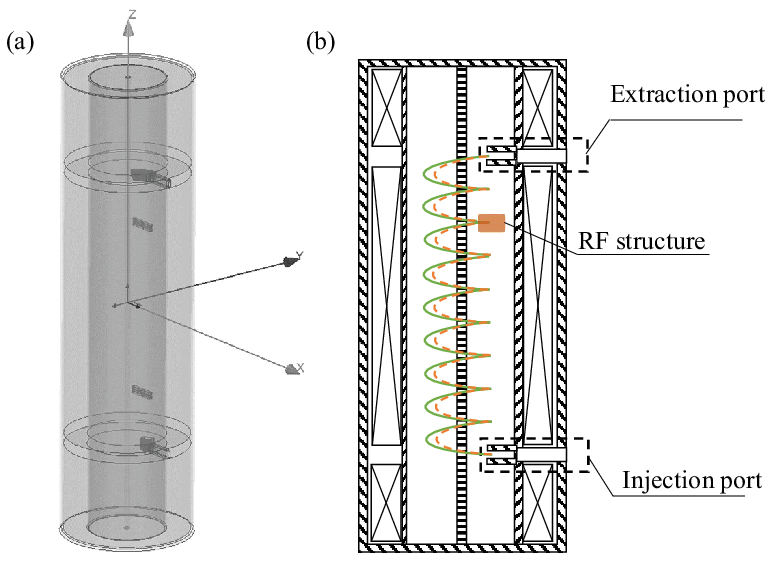}
		\caption{(a) Compression installation (b) The front view profile of the compression installation. It comprises a primary solenoid, injection and extraction structures, several RF Structures.}
		\label{FIG.2}
	\end{figure}

	As depicted in FIG. \ref{FIG.2}, The device consists of a main coil, injection and extraction magnets, and local RF cavities.
	
	The main coil comprises three segments with independently adjustable current, generating a uniformly distributed axial magnetic field within the cylindrical cavity. Injection and extraction ports for the electron beam are positioned at the interfaces between the three coil segments. The injection and extraction structure is composed of a combination of permanent magnets and iron, ensuring that electrons, upon injection into the cylindrical cavity from the exterior, can execute a coaxial spiraling motion within the cavity and ultimately exit the set-up through the extraction structure. The local RF cavity is positioned along the electron trajectory for the longitudinal focusing of the electron beam bunch.
	
	\begin{figure}[htbp]
		\includegraphics[width=8.5cm]{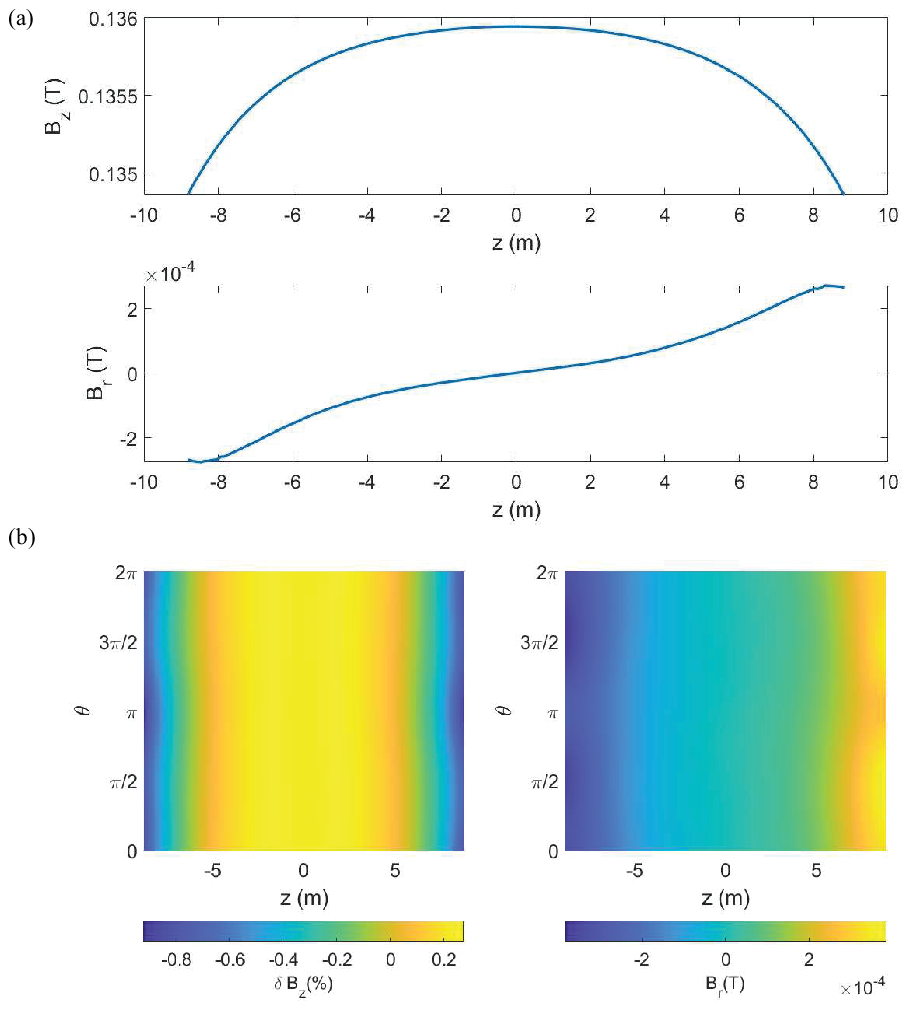}
		\caption{Uniformilty of magnetic field in the cavity. (a) Magnetic field component $B_z$ and $B_r$ along the z-direction at $R = 1.5 \; m$ (b) Uniformity of the axial magnetic field $B_z$ on cylindrical surface $R = 1.5 \; m$, where $\delta B_z = \frac{B_z-\overline{B_z}}{\overline{B_z}}$ (c) Distribution of radial magnetic field $B_r$.  The field is obtained through a CST Studio simulation.}
		\label{FIG.3}.
	\end{figure}

	The coils generats a nearly uniform magnetic field of approximately 0.14 T within its interior volume. Dimensions of the installation are an inner diameter of 4.8 m, an outer diameter of 6 m, and a height of approximately 29 meters. The magnetic field distribution of the solenoid coil and the injection/extraction structure is illustrated in FIG.\ref{FIG.3}.
	
	In the upcoming sections, we will use dynamic simulation to illustrate the transmission properties of this set-up and explore two distinct compression modes: high-frequency pulse compression for generating THz electron beams and continuous beam compression for producing high-energy intense electron beams.

	\section{Simulation}

\begin{figure}[htbp]
		\includegraphics[width=8.5cm]{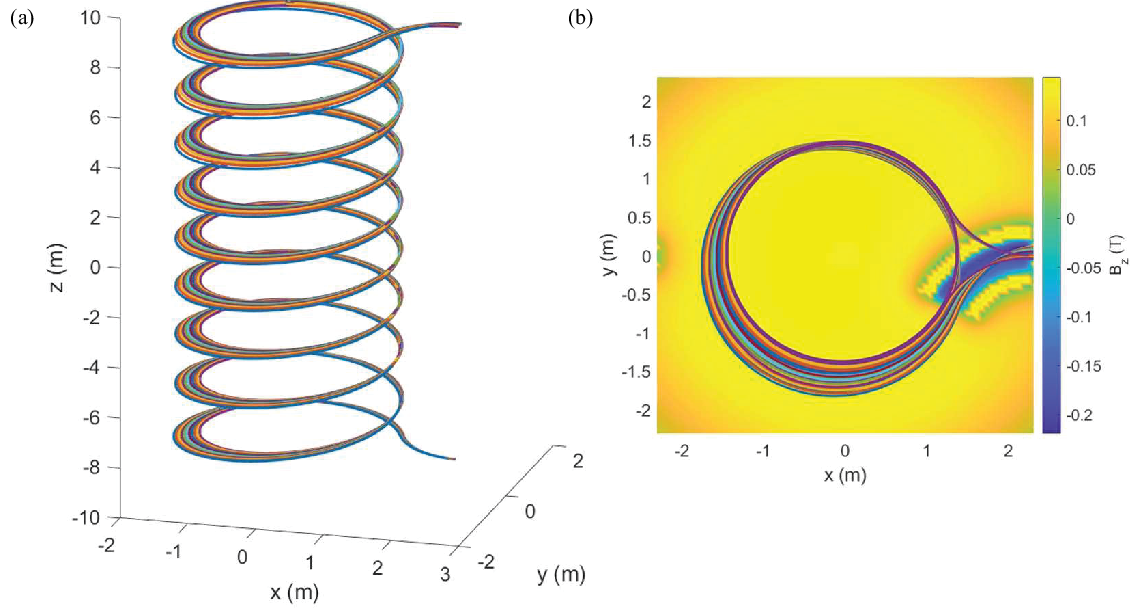}
		\caption{Trajectories of electrons with energy dispersion of of ±5\% for the reference energy of 6.5 MeV. (a) 3-dim trajectories (2) Projection on x-y plane.}
		\label{FIG.4}.
	\end{figure}

	In simulation, a reference electron energy of 60 MeV is selected. FIG.\ref{FIG.4} depicts electron trajectories with an energy dispersion of ±5\% for the reference energy of 60 MeV.
	
	The trajectory radius of the reference electron $R \approx 1.5$ m, and the revolution period $T_c \approx 30.8$ ns. The electrons undergo 9 cycles of spiral motion from injection to extraction in this transportation system. According to function \ref{eq1}, the ideal momentum compaction factor $R_{56}=83.2$ m.

	However, due to the existence of fringe fields at the connections to the injection and extraction structures, and the non-ideal uniform field distribution inside the solenoid cavity, the transmission system is nonlinear. The effect of the transmission system cannot be represented by a simple transfer matrix.
	
		\begin{figure}[htbp]
		\includegraphics[width=6cm]{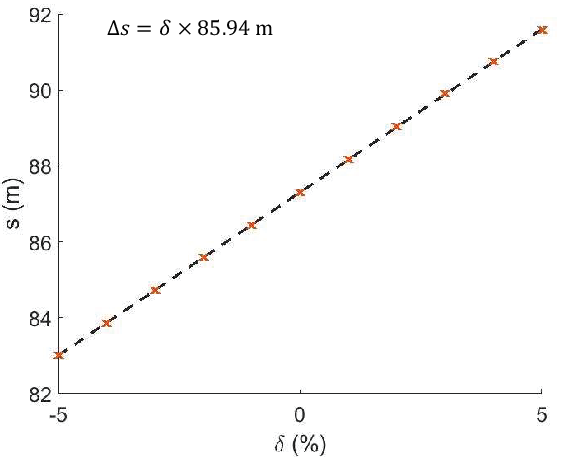}
		\caption{The latency of the transmission system and the energy of the electrons satisfy a linear relationship. $\delta = \frac{\Delta E}{E_0}$.}
		\label{FIG.5}.
	\end{figure}
	
	According to the dynamic simulation, the actual momentum compaction factor $R_{56}=85.94$ m, which is slightly greater than the estimated value under ideal homogeneous field conditions. Fortunately, the distance latency and the energy spread of the electrons still adhere to an ideal linear relationship. This ensures that the system can achieve beam compression as we intended.
	
\begin{figure}[htbp]
		\includegraphics[width=8.3cm]{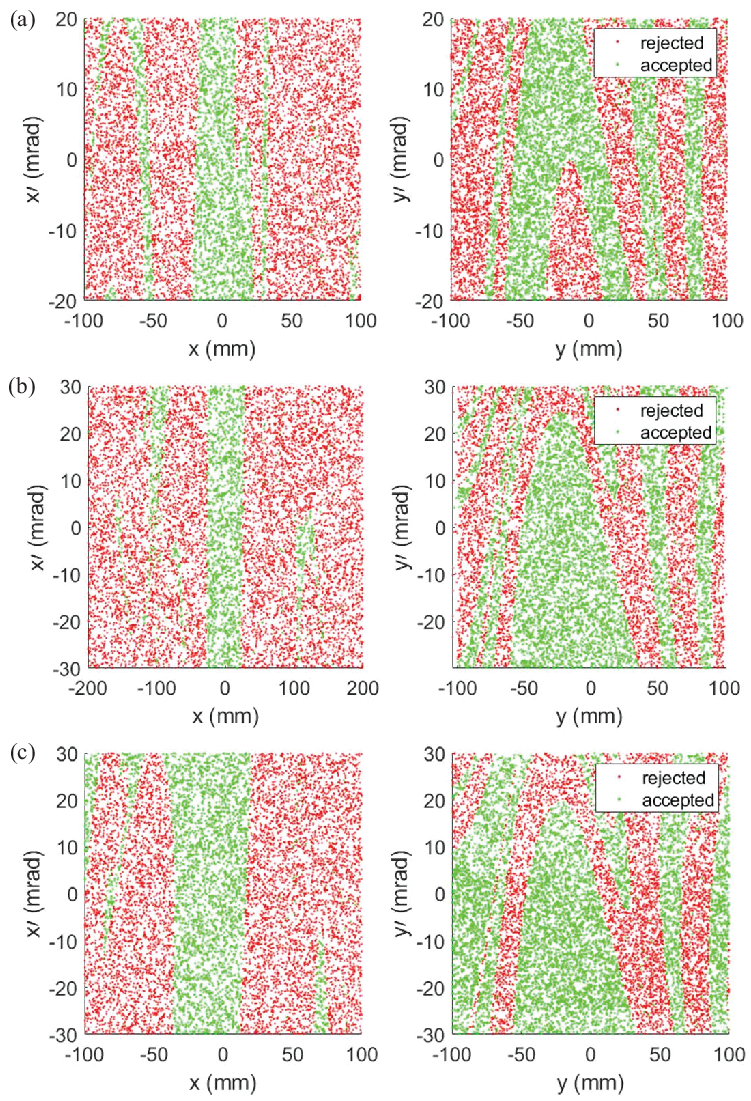}
		\caption{Transverse acceptance for electrons with energy of (a) 58 MeV, (b) 60 MeV and (c) 62 MeV. 'Accepted' indicates that after the spiral movement, electrons can be drawn from the extraction structure of the compression device, while 'rejected' suggests the opposite.}
		\label{FIG.6}.
	\end{figure}

	As described in part of the set-up, the electron beam undergoes a spiral motion over a length of 80 meters, passing through only one local RF cavity. No lattice is set up at other positions to control or constrain it. Hence, this system has a very large aperture. FIG. \ref{FIG.6} presents the acceptance of the transmission system for electrons with energies of 58, 60 and 62 MeV. The electron's velocity is almost along the angular direction of the laboratory column coordinate system, therefore the x-direction in the phase space almost coincides with the radial direction of the laboratory coordinates, while the y-direction essentially follows the axial direction.
		
	The intersection ellipse of the transverse acceptance phase diagrams in the energy range of $60 MeV \pm 5\%$ is marked in FIG. \ref{FIG.6}, with a transverse geometric acceptance $\epsilon \approx 10 \; \pi \; mm \cdot mrad$ .

	It is worth mentioning that this compression device is non-dispersive. The longitudinal energy spread of the electron beam will eventually manifest as a larger transverse emittance of the beam. However, due to the large aperture, it does not affect the extraction of the beam after compression.

	\subsection{Compression for sub-THz beam generation}
	
	The primary objective of our design is to increase the frequency of the beam generated by the RF accelerator to the THz range. Therefore, we chose a beam with a frequency of 3 GHz for compression. 
	
\begin{figure}[htbp]
		\includegraphics[width=8.5cm]{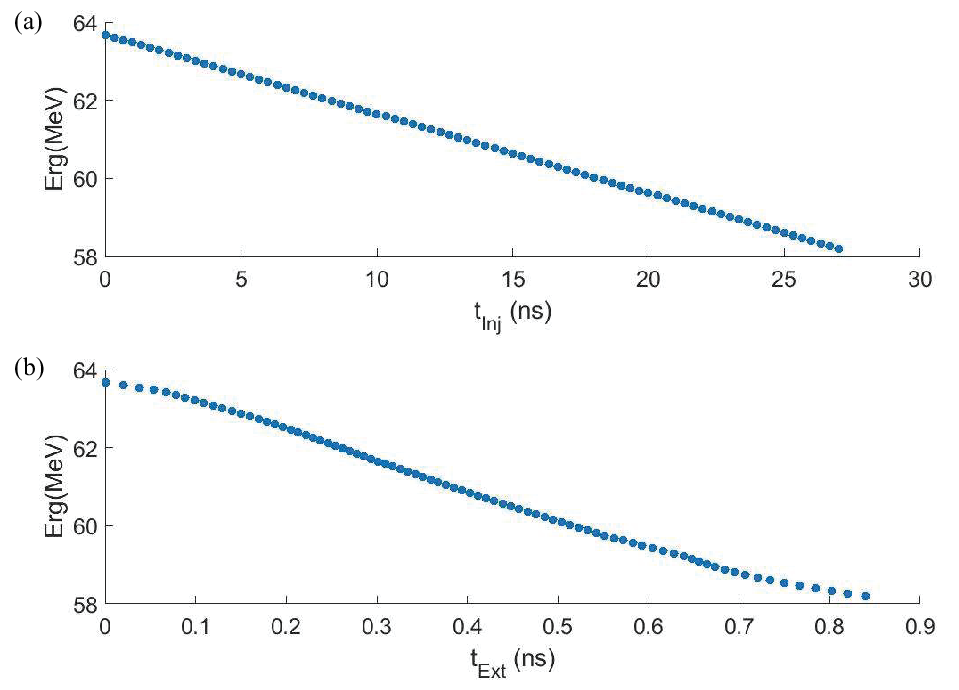}
		\caption{The energy-time distribution before and after beam compression using the point charge model.}
		\label{FIG.7}.
	\end{figure}

	Beginning with a basic model, we substituted point charges for micro-pulses, depicting the energy-time distribution of the beam pre and post compression in FIG. \ref{FIG.7}. The initial pulse width of the electron beam is approximately 27 ns, with an energy range of 58 to 63 MeV. After compression, the length of the macro-pulse has been compressed to approximately 0.85 ns.
	
	While the linear relationship between time delay and energy spread in the transmission system is relatively ideal, upon closer examination at a 10 ps scale, uneven spacing between the micro-bunches post-compression persists.Moreover, when dealing with micro-bunches instead of point charges, it is essential to factor in the pulse width of the micro-pulses. Our objective is not solely compressing a lengthy macro-pulse but also maintaining a periodic micro-pulse structure post-compression.
	
	To tackle these challenges, we employed two solutions.
	
	Firstly, prior to injecting the beam into the compression structure, we considered not only the linear energy-time modulation of the macro-pulse as a whole but also the energy modulation of each micro-pulse. Secondly, we positioned a local RF structure along the beam trajectory for longitudinal focusing of the micro-bunches and fine-tuning the bunch spacing through negative feedback control.
	
\begin{figure}[htbp]
		\includegraphics[width=8.5cm]{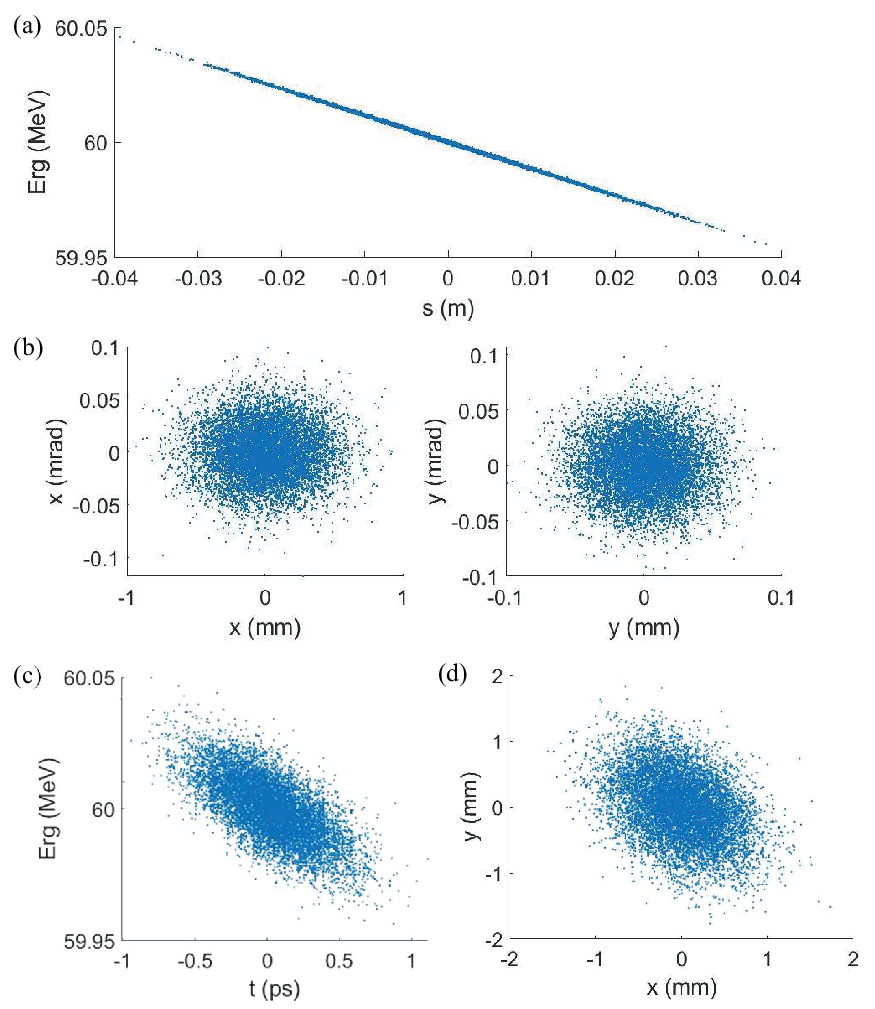}
		\caption{The (a) longitudinal and (b) transverse distribution of a micro-bunch before injection.The (c)  longitudinal and (d) transverse distribution of a micro-bunch after extraction.}
		\label{FIG.8}.
	\end{figure}

	The longitudinal distribution of micro-bunches before compression is depicted in Figure 8 (a).  Each micro-bunch underwent energy pre-modulation with a slope of $\frac{s}{\delta}=85$ m. Each micro-bunch had a 1‰ energy spread.  The transverse distribution, as shown in Figure B, exhibits an initial geometric emittance of 0.1 mm mrad in the horizontal direction and 0.01 mm mrad in the vertical direction. Without considering space charge forces, the pulse width at the extraction of each micro-bunch is  less than 10 ps (FIG. \ref{FIG.8} (c)), meeting the condition where adjacent micro-bunches do not overlap.
	
	However, the presence of space charge forces exacerbates the situation. When the charge of the micro-bunches reaches 50 pC, after undergoing such long-distance free drift, the longitudinal size of the micro-bunches expands and the waveform of the extracted beam micro-pulses becomes blurred. Furthermore, with non-uniform bunch spacing, the form factor of the current signal post-compression at a frequency of 0.1 THz $f <0.1$.
	
	Thus we proceeded with a further solution. A pair of RF electrodes is placed at the position shown in Figure 9(a), creating a local RF field to provide longitudinal focusing for the passing micro-bunches.
	
	As mentioned earlier, this RF field needs to simultaneously serve the purposes of micro-bunch focusing and bunch spacing adjustment. However, the RF field amplitudes required for these two tuning objectives are not consistent. Therefore, we utilized a dual-frequency hybrid RF setup in conjunction with adjusting the initial longitudinal modulation slope of the micro-bunches to accommodate both goals simultaneously.

	\begin{figure}[htbp]
		\includegraphics[width=8.5cm]{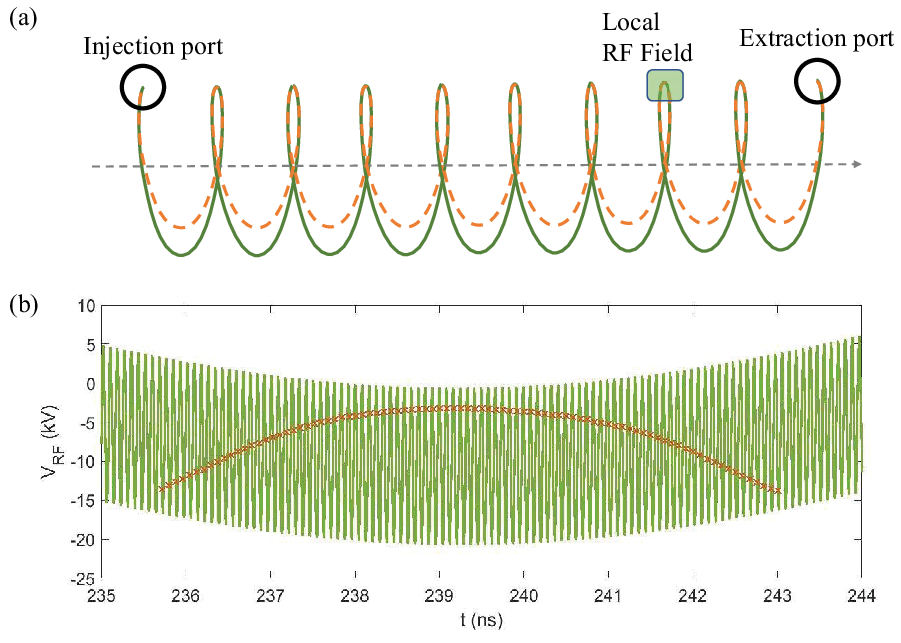}
		\caption{Longitudinal distribution of the bunch train (a) before and (after) the compression. (c) Arrival time and the (d) corresponding Phase of each micro bunch at RF1.}
		\label{FIG.9}.
	\end{figure}
	
	The two RF signal frequencies are $f_1=12$ GHz and $f_2=40$ MHz, with amplitudes of $V_1=10$ kV and $V_2=10.7$ kV respectively. The effective working length of the local RF field is 10cm. The synthesized signal waveform is shown in FIG \ref{FIG.9} (b). The arrival time of each micro bunch at the RF Field and the corresponding accelerating voltage are also denoted. 
	
	We use the code General Particle Tracer to simulate the compression process under large charge conditions. 
	
	\begin{figure}[htbp]
	\includegraphics[width=8.5cm]{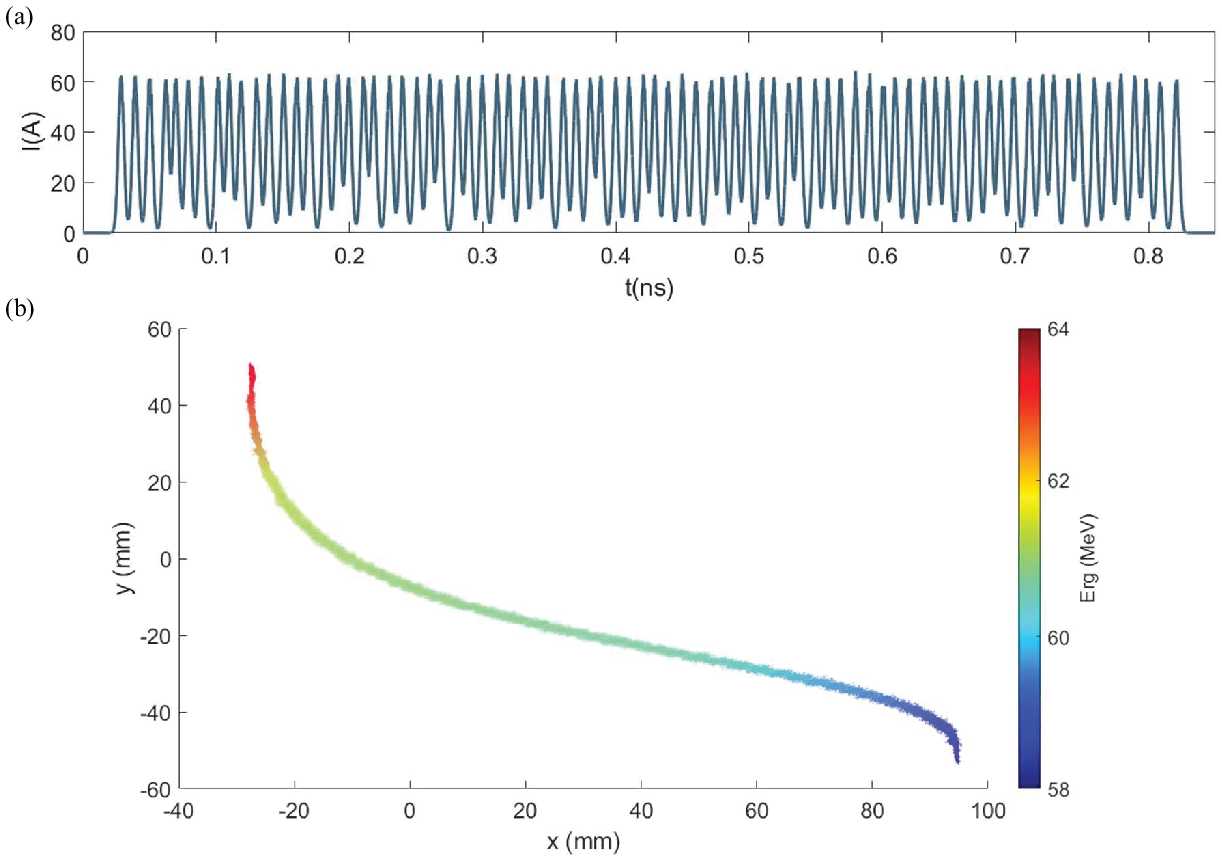}
	\caption{(a)The current profile and (b) transverse distribution of electron beam  after compression.}
	\label{FIG.10}.
	\end{figure}

	The initial transverse and longitudinal distributions of the microbunches before compression remain as depicted in FIG.\ref{FIG.8} (a-b). Through calculations, the maximum allowable charge for each microbunch during this transmission process is 0.3 nC,  to prevent overlap and confusion between adjacent microbunches. Thus, we obtain a relativistic electron pulse train at the extraction port with a frequency of 0.1 THz, a macro-pulse width of  approximately 0.8 ns, a total charge of 24 nC, and energy of 1.44 J. The current waveform after compression is illustrated in FIG.\ref{FIG.10} (a), with a shape factor of $F=0.35$ at the 0.1 THz frequency. The peak current is 60 A, with a peak power of 3.6 GW, while the average current is 34 A, and the average power is 2.04 GW

	However, as this transmission system is non-dispersive in the transverse direction, the $10\%$ beam divergence leads to a notably large transverse size at the extraction port (as depicted in FIG.\ref{FIG.10} (b) ). While our method offers a possibility for producing ultra-intense THz beams using RF accelerators, resolving these issues will require further research for practical THz applications.

	\subsection{Continues beam compression}
	
	The significant transverse emittance presents a challenging issue in THz-related application contexts. However, in applications such as white neutron sources where the electron beam directly impacts the target, its emittance becomes less critical\cite{GELINA,danon2016recent}. Consequently, we have explored this compression method for scenarios involving continuous beam compression. It resembles GELINA's magnetic compression device\cite{GELINA2}. It employs a 3-meter diameter, 50-ton magnet to compress a beam with a reference energy of 100 MeV and a 50\% energy spread from 10 ns to 1 ns. Our device achieves compression from 30 ns to 1 ns for a beam with a reference energy of 60MeV and a 10\% energy spread.
	
	This compression mode is fundamentally a power compression technique. In the field of particle accelerators, common media used for power compression include electricity, microwaves, and lasers. Relativistic electron beams offer advantages as energy carriers. For instance, it is not constrained by device breakdown voltages, boasts an exceptionally high energy storage capacity ($P \propto \gamma^3$), and achieves nearly 100\% compression efficiency.
	
\begin{figure}[htbp]
		\includegraphics[width=8.5cm]{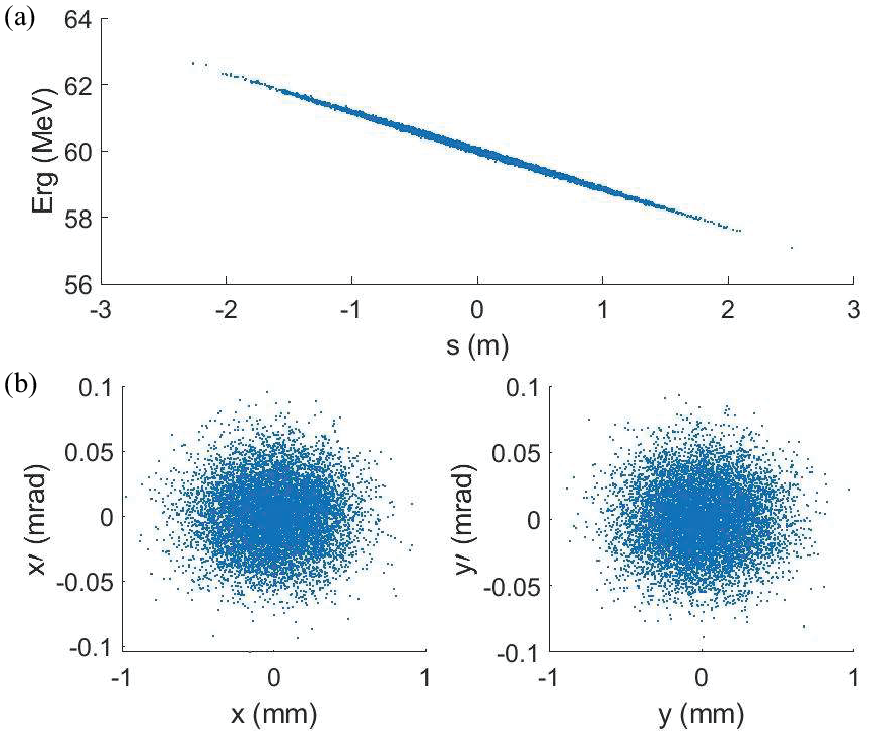}
		\caption{(a) The longitudinal and (b) transverse distribution of the beam before compression.}
		\label{FIG.12}.
	\end{figure}

	In simulation with General Particle Tracer and MATLAB, the transverse and longitudinal distributions of the beam before compression are shown in FIG. \ref{FIG.12}, with an initial geometric emittance of 0.1 mm$\cdot$mrad.
	
\begin{figure}[htbp]
		\includegraphics[width=8.5cm]{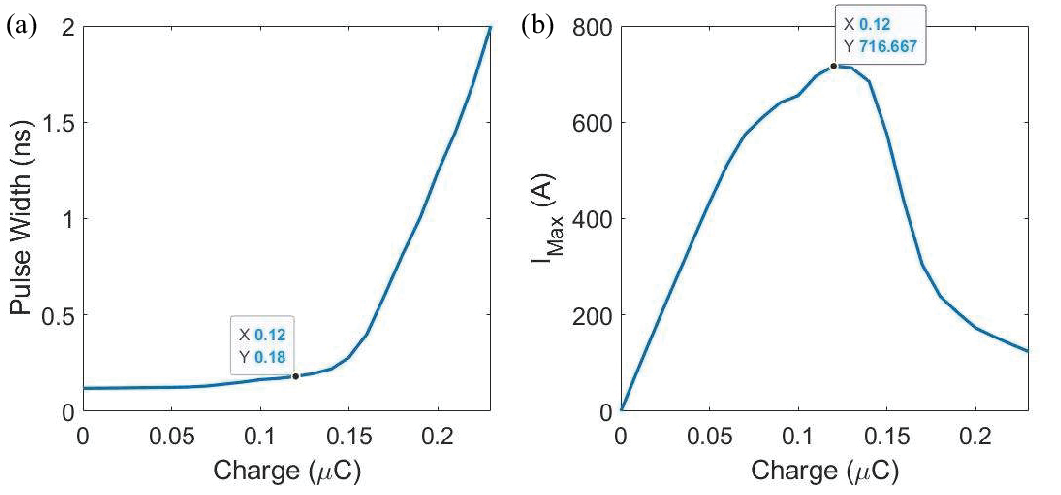}
		\caption{The relationship between the (a) half-width at half-maximum (HWHM),  (b) peak current of the beam after compression, with the total charge of the bunch.}
		\label{FIG.13}.
	\end{figure}

	In this continuous mode, the primary objective is to maximize the post-compression average beam intensity without any particle loss. In the simulations, we scanned the charge of the beam $Q$, and established the relationship between the peak current of extracted beam $I_{max}$ and charge quantity $Q$ (FIG. \ref{FIG.13}). For $Q < 0.12 \;  \mu $C, the peak current increased with the total charge quantity. However, when $Q > 0.12 \;  \mu $C, the space charge force caused longitudinal expansion of the beam after compression, leading to a decrease in peak current as the total charge increased. When $Q > 0.2 \;  \mu $C, particle loss began due to the limited aperture of the device. Therefore, at a charge quantity of $Q =0.12 \;  \mu $C, we obtained the maximum peak current of the extracted beam. $I_{max} = 716.7$ A, $P_{max} = 43.0$ GW.
	
\begin{figure}[htbp]
		\includegraphics[width=8.5cm]{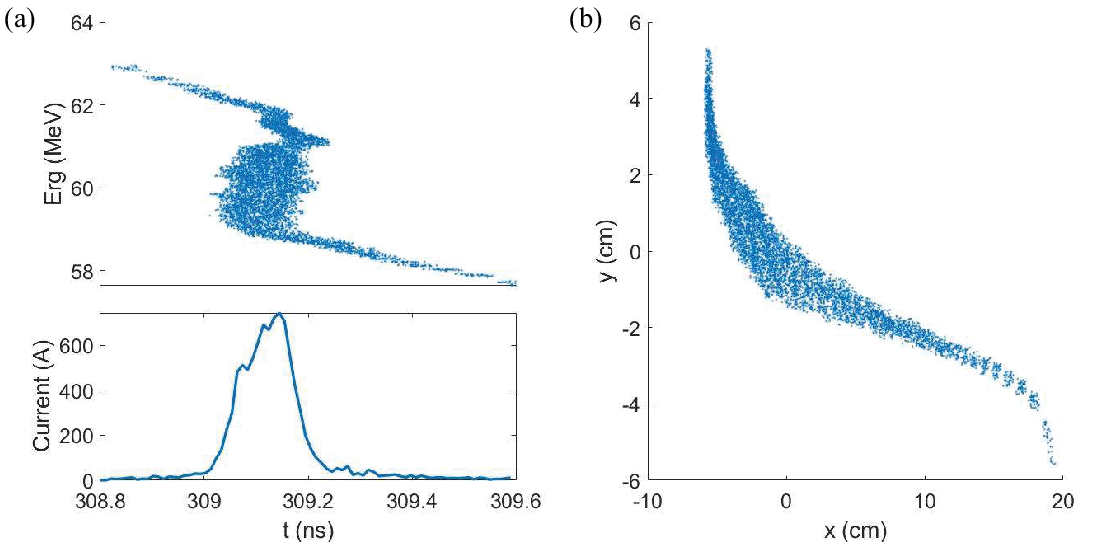}
		\caption{(a) The longitudinal and (b) Transverse distribution of the extracted beam.}
		\label{FIG.14}.
	\end{figure}

	The distribution of the beam in all dimensions after compression is shown in FIG. \ref{FIG.14}. For applications other than target hitting, additional focusing considerations are necessary before deployment.
	
	\section{conclusion}
	Our research introduces a intriguing method for relativistic electron beam macro-pulse compression, utilizing the dispersion effect of electrons in a uniform magnetic field to modulate beam density. The nearly unrestricted spiral motion of electrons within the coil cavity offers a larger dynamic aperture at a lower processing cost compared to intricate ring structures, albeit resulting in a larger transverse size of the compressed beam.
	
	Drawing from our previous research, similar devices synthesized electron pulse trains through a different operational mode\cite{li2024longitudinal}, effectively controlling the electron's transverse distribution with dipole deflecting and quadrupole focusing magnets along the trajectory. Our ongoing efforts aim to optimize the beam emittance of the compressed beam through this approach.
	
	This research on compression methods introduces a new potential for generating THz electron beams with ultra-high intensity. While the beam quality and the large device volume remain significant challenges. We are committed to further model refinements and the advancement of validation experiments to bridge this gap.
	
	\section{acknowledgments}
	
	The authors would like to thank Alex Chao for useful discussions and suggestions. His proposition that our device could be utilized to generate THz electron beams sparked the intriguing idea that led to the further exploration detailed in this paper.
	
	\bibliography{ref}

\end{document}